\documentclass[showpacs,preprintnumbers,amsmath,amssymb,prl,aps,preprint]{revtex4}

\usepackage{epsfig}
\usepackage{graphicx}
\usepackage{dcolumn}
\usepackage{bm}

\begin{document}

\title{Manifestation of lattice distortions in the O 1$s$ spectra in Ca$_{1-x}$Sr$_x$RuO$_3$}

\author{Ravi Shankar Singh and Kalobaran Maiti
\footnote{Corresponding author: e-mail: kbmaiti@tifr.res.in~~~~~
Fax: +91 22 2280 4610}}

\affiliation{Department of Condensed Matter Physics and Materials
Science, Tata Institute of Fundamental Research, Homi Bhabha Road,
Colaba, Mumbai - 400 005, INDIA}

\date{\today}

\begin{abstract}

We investigate the effect of temperature on the electronic and
crystal structures of Ca$_{1-x}$Sr$_x$RuO$_3$ via the evolution of
O 1$s$ core level spectra as a function of temperature and
composition, $x$. O 1$s$ spectra in SrRuO$_3$ exhibit a dominant
sharp feature at all the temperatures with a small trace of
impurity feature at higher binding energies. The spectra in Ca
doped samples, however, exhibit two distinct features. Analysis of
the experimental spectral functions and the band structure results
suggest that different Madelung potential at the two types of
oxygen sites in the orthorhombically distorted structure leads to
such splitting in the O 1$s$ spectra. Interestingly, the energy
separation of these two features becomes smaller at low
temperatures in the Ca dominated samples concomitant to the
observation of non-Fermi liquid behavior in their bulk properties.
Such temperature evolution, thus, indicates a direct connection of
the lattice degrees of freedom with the electronic properties of these compounds.\\


\end{abstract}


\maketitle

\section{Introduction}

Research in transition metal oxides has seen an explosive growth
during last few decades due to discovery of many exotic properties
such as high temperature superconductivity, giant
magnetoresistance, insulator to metal transitions, quantum
confinement effects {\it etc}. It is believed that all these novel
material properties are essentially determined by the electronic
states corresponding to the transition metal and oxygens forming
the valence band (the highest occupied band). Numerous studies
have been carried out to achieve microscopic understanding
focussing primarily on the role of electron correlations in these
material properties. The difficulty in such studies is that a
large number of parameters such as electron correlation,
electron-lattice interactions, charge carrier concentration,
disorder {\it etc.} play crucial roles and it is virtually
impossible to study the effect of a single parameter independently
in a real system. In addition, significant mixing of the O 2$p$
and transition metal $d$ states leads to further complexity of the
problem.

Recently, there is a growing realization that the interaction of
electrons and lattice vibrations is important to determine various
exotic material properties such as colossal magnetoresistance
\cite{millis}, pseudogap phase in high-temperature superconductors
\cite{mannella}, strange metallic behavior\cite{ruth-nfl} {\it
etc}. Since, oxygens in these materials have valency close to (2-)
(electronic configuration, 2$s^2$2$p^6$), the O 2$p$ levels are
essentially filled. Thus, final state effects such as the
correlation effects, core hole screening due to charge transfer
from transition metals {\it etc.} will be negligible and the
evolution of the O 1$s$ core levels are expected to efficiently
manifest primarily the lattice effects.

In the present study, we investigate the role of electron-lattice
interactions in the ground state properties of
Ca$_{1-x}$Sr$_x$RuO$_3$ via the evolution of O 1$s$ core levels as
a function of temperature and composition. SrRuO$_3$ is a
perovskite compound and exhibits ferromagnetic long range order
below $\sim$~160~K \cite{rss,cao}. However, no magnetic long-range
order is observed in isostructural, CaRuO$_3$ down to the lowest
temperature studied \cite{rss,cao,nfl,klein}. Various
investigations predict a non-Fermi liquid ground state in
CaRuO$_3$ in contrast to the Fermi-liquid behavior observed in
SrRuO$_3$ \cite{nfl,klein}. Recently, it is shown that the
correlation effects are significantly weak in both these compounds
\cite{ruth-corr}. Band structure results using full potential
augmented plane wave method suggest that Ca-O/Sr-O covalency plays
the key role in determining the electronic as well as crystal
structure in these systems, and the absence of long range order in
CaRuO$_3$ has been attributed to the smaller Ru-O-Ru angle of
150$^\circ$ compared to 165$^\circ$ in SrRuO$_3$ \cite{ruth-band}.

While all these studies throw some light on the electronic and
magnetic properties of these systems, transition from Fermi liquid
to non-Fermi liquid behavior is still a puzzle. A recent
photoemission study using high-energy resolution reveals signature
of particle-hole asymmetry and predict strong influence of phonons
in the electronic excitation spectra in CaRuO$_3$ in contrast to
the case of SrRuO$_3$. It is now timely to investigate the lattice
effects independently to achieve microscopic understanding of the
electronic properties in these systems. We observe that multiple
features appear in the oxygen 1$s$ spectra due to orthorhombic
distortion of the crystal structure in CaRuO$_3$, while close to
cubic structure of SrRuO$_3$ leads to a single feature in the
spectra. The photoemission results reveal significant temperature
induced modification of the crystal lattice in Ca dominated
compositions, while SrRuO$_3$ appears to be very similar at all
the temperatures studied.

\section{Experimental}

High quality polycrystalline samples (large grain size achieved by
long sintering at the preparation temperature) were prepared by
solid state reaction method using ultra-high purity ingredients
and characterized by $x$-ray diffraction (XRD) patterns and
magnetic measurements as described elsewhere \cite{rss,ruth-corr}.
Sharp XRD patterns reveal pure GdFeO$_3$ structure at all the
compositions studied with the lattice constants similar to those
observed for single crystalline samples \cite{cao}. Photoemission
measurements were carried out in an ultra-high vacuum chamber
(base pressure lower than 5$\times$10$^{-11}$~torr) using SES2002
Scienta analyzer, and Al $K\alpha$ $x$-rays from  monochromatic
and twin sources. The energy resolutions were fixed at 0.3~eV and
0.9~eV for the measurements using monochromatic and twin sources,
respectively. Sample surface was cleaned by {\it in situ}
scraping. Reproducibility and cleanliness of the sample surface
was confirmed after each trial of scraping.

\section{Results and discussions}

\begin{figure}
\vspace{-4ex}
 \centerline{\epsfysize=4.5in \epsffile{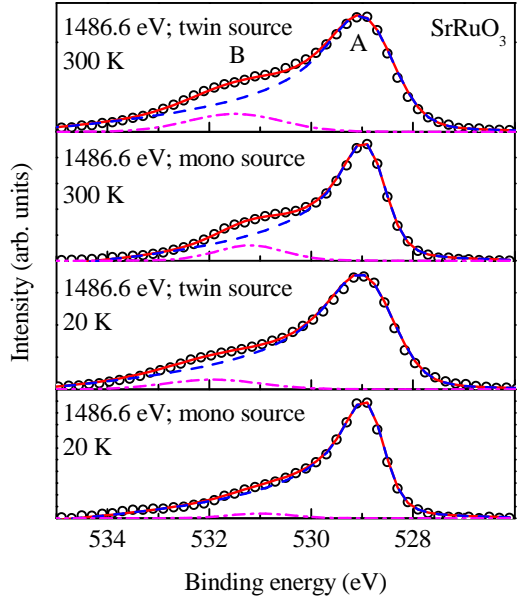}}
\vspace{-16ex}
 \caption{O 1$s$ core level spectra of SrRuO$_3$ at different temperatures.}
 \vspace{-2ex}
\end{figure}

In Fig.~1, we show the O 1$s$ core level spectra of SrRuO$_3$ at
different temperatures and photon sources. The Al $K\alpha$ twin
source spectrum at 300~K exhibit two distinct features. The
feature A appears at about 528.8~eV binding energy and has the
largest intensity. The second feature B is significantly weak
compared to A and appears around 531.5~$\pm$~0.2 eV. Since, O 2$p$
levels are almost completely filled, the feature B cannot be
attributed to photoemission signals due to different degrees of
core hole screening in the photoemission final states as often
observed in the core level spectra corresponding to transition
metals and/or rare earths. Thus, this features is often used as a
measure of the additional oxygens and/or foreign oxides adsorbed
on the sample surface. Scraping the sample surface often leads to
decrease in intensity of this feature confirming the above
predictions. In the present case, the intensity of feature B could
not be reduced further by repeated scrapings. In order to estimate
the contribution of feature B in the total spectrum, we fit the
experimental spectra by two Doniach-\v{S}unji\'{c}
lineshapes\cite{don-sun} as shown by dashed and dot-dashed lines
in the figure. The simulated spectral function passing through the
experimental data points represent a good fit. It is clear that
the asymmetry of the feature B is significantly small compared to
that for the feature A. The intensity of B is about 10\% of the
total intensity.

\begin{figure}
\vspace{-4ex}
 \centerline{\epsfysize=4.5in \epsffile{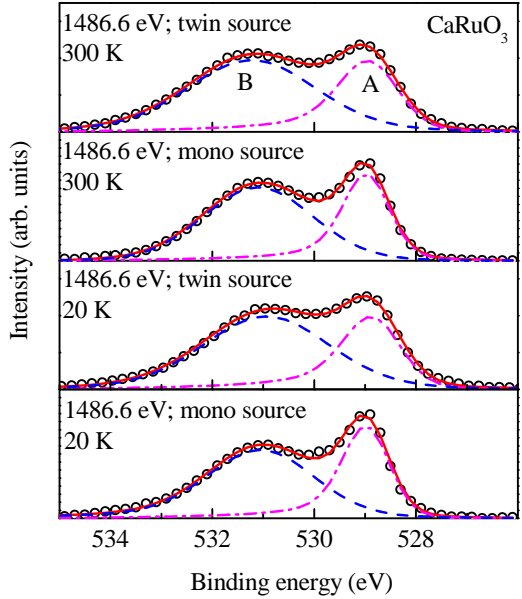}}
\vspace{-16ex}
 \caption{O 1$s$ core level spectra of CaRuO$_3$ at different temperatures.}
 \vspace{-2ex}
\end{figure}

The second panel of Fig. 1 shows the O 1$s$ signals from SrRuO$_3$
using monochromatic $x$-ray source, which has significantly
improved energy resolution (0.3~eV). It is clear that the spectral
features A and B are narrower and somewhat better defined. The
relative intensity of the feature B is found to be about 9\% of
the total intensity similar to that observed in the case of twin
source spectrum. The intensity of feature B reduces significantly
with the decrease in temperature to 20 K (see third and fourth
panel of Fig.~1) corresponding to un-monochromatized (6\% of total
intensity) and monochromatized source ($<$~3\% of total
intensity), respectively. It is to note here that the intensity of
feature B is always found to be slightly higher in the twin source
spectra compared to that in the monochromatic source spectra
presumably due to spectral weight redistribution by the larger
width of the incident light. These results establish that {\it it
is possible to generate clean sample surface by scraping high
quality polycrystalline samples}.

The O 1$s$ spectra of CaRuO$_3$ are shown in Fig.~2. The two
features A and B appear around 528.7~eV and 531.4~eV binding
energies, respectively. Surprisingly, the feature B is always
found to be significantly broad and intense with an integrated
intensity higher than feature A. In order to investigate, if this
large intensity appears due to the impurities at the sample
surface, grain boundaries and/or due to multiple phases of
CaRuO$_3$, we have prepared 3 sets of samples. The sharp features
in the $x$-ray diffraction patterns and absence of any additional
peak suggest high quality and single phase of the samples. In
addition, the magnetic measurements exhibit magnetic moment of
3~$\mu_B$/fu in the paramagnetic phase of CaRuO$_3$
(2.7~$\mu_B$/fu in SrRuO$_3$), which is close to their spin-only
value of 2.83~$\mu_B$ for $t_{2g\uparrow}^3 t_{2g\downarrow}^1$
configurations at Ru sites \cite{rss,cao}. This suggests that the
two peak structure in O 1$s$ spectra might be intrinsic to this
sample.

\begin{figure}
\vspace{-4ex}
 \centerline{\epsfysize=4.5in \epsffile{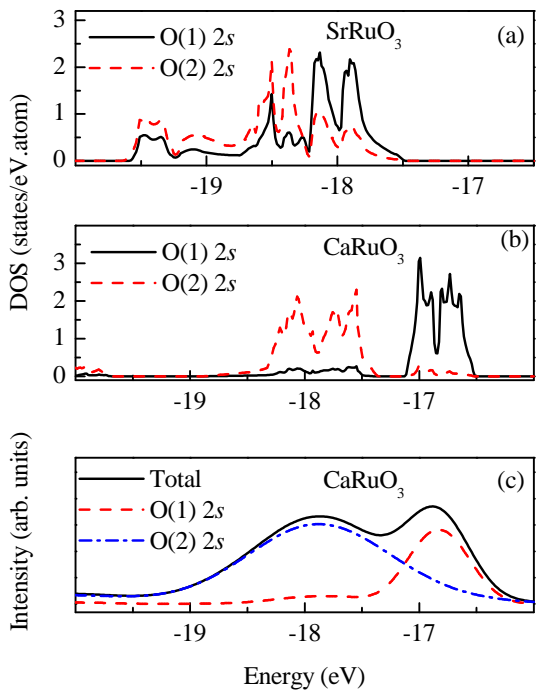}}
\vspace{-8ex}
 \caption{The density of states of O 2$s$ core levels in
 (a) SrRuO$_3$ and (b)CaRuO$_3$ obtained by band structure calculations.
 The calculated results exhibit well separated O(1) and O(2) contributions
 in CaRuO$_3$ compared to that in SrRuO$_3$. (c) The O 2$s$ density of states
 in CaRuO$_3$ are broadened by a Gaussian to show the lineshape of the total
 spectral function.}
 \vspace{-2ex}
\end{figure}

In order to investigate the origin of these two features, we have
calculated the O 2$s$ core level density of states using
state-of-the-art full potential linearized augmented plane wave
(FPLAPW) method (Wien2K software \cite{wien}). There are two types
of oxygens present in the structure. If O(1) represent the apical
oxygens of the RuO$_6$ octahedra and O(2) represent those in the
basal plane, there are one O(1) and two O(2) atoms in one formula
unit. It is important to note here that the electron electron
Coulomb interactions cannot be treated exactly in these
calculations. Thus, binding energy of the core levels is often
underestimated in the calculations compared to the experimental
results. However, the effect due to Madelung potential can be
captured reasonably well in these calculations. The details of the
method of calculation is described elsewhere \cite{ruth-band}.

The calculated results for SrRuO$_3$ and CaRuO$_3$ are shown in
Fig.~3(a) and 3(b), respectively. The O 2$s$ contributions in
SrRuO$_3$ appear very close to each other in Fig.~3(a). However,
the density of states in CaRuO$_3$ is significantly different. The
separation between O(1) 2$s$ and O(2) 2$s$ contributions is more
than 1~eV. If we broaden the two features with a Gaussian with a
slightly different width, the lineshape of the resultant spectral
function shown in Fig.~3(c) is remarkably similar to that observed
in Fig.~2. In order to investigate the intensity ratio of the two
features in the experimental spectra, we fit the experimental
spectra in Fig.~2 by a set to two Doniach-\v{S}unji\'{c}
lineshapes as done in the case of SrRuO$_3$. Interestingly, the
intensity of the feature B is found to be about 1.8~$\pm$~0.2
times of the intensity of the feature A. These results, thus,
reveal that the two peak structure in the O 1$s$ spectra in Fig.~2
is intrinsic and can be attributed to different Madelung potential
at different oxygen sites.

Decrease in temperature leads to a shift in the peak position of
feature B by about 0.2~eV towards lower binding energies. The
relative intensity of the features remain almost similar at low
temperatures. Notably, the charge transfer satellite in Ca 2$p$
core level spectra in CaRuO$_3$ also moves towards lower binding
energies with the decrease in temperature \cite{ruth-epl}.
However, Sr 3$d$ core level spectra does not exhibit such effect
in SrRuO$_3$. This clearly indicates that the decrease in
temperature leads to a significant modification in the local
structure involving Ca and O sites in CaRuO$_3$.

\begin{figure}
\vspace{-4ex}
 \centerline{\epsfysize=4.5in \epsffile{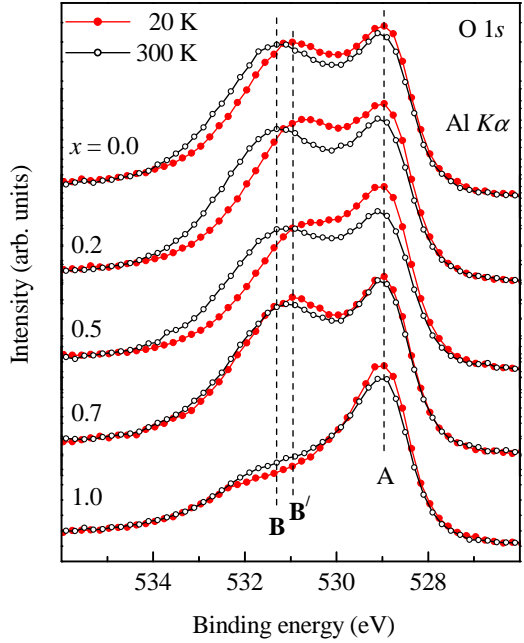}}
\vspace{-12ex}
 \caption{O 1$s$ core level spectra of Ca$_{1-x}$Sr$_x$RuO$_3$ for
 different values of $x$ at 300 K and 20 K. All the spectra are collected using
 Al $K\alpha$ un-monochromatized source.}
 \vspace{-2ex}
\end{figure}

In order to investigate this effect across the whole composition
range, we plot the O 1$s$ spectra in Ca$_{1-x}$Sr$_x$RuO$_3$ for
different values of $x$ in Fig.~4. All the spectra are normalized
by total integrated area under the curve. Two effects are clearly
visible in the figure. Firstly, the feature B is significantly
strong even in the 70\% Sr substituted sample. Thus, a small
amount of Ca in the crystal lattice appears to introduce a
significant structural distortion leading to two intense features
as observed in CaRuO$_3$. Secondly, the feature B at 300 K shifts
by about 0.2~eV to B$^\prime$ in the 20 K spectra for all $x$
values upto $x$ = 0.5. This shift becomes very small for $x$ = 0.7
and almost not significant for $x$ = 1.0. This suggests that the
temperature induced local distortion primarily appears in the Ca
dominated compositions. Interestingly, the non-Fermi liquid
behavior is also observed in the samples with similar compositions
\cite{nfl}. This study thus, provides an evidence of a direct
relationship of the non-Fermi liquid behavior with the structural
changes.

In summary, we investigate the evolution of O 1$s$ core level
spectra as a function of temperature and composition in
Ca$_{1-x}$Sr$_x$RuO$_3$ for various values of $x$. The spectra in
SrRuO$_3$ shows that high quality polycrystalline samples can be
cleaned efficiently by {\it in situ} scraping. The scraped surface
at low temperature appears to be almost free of any impurity. This
is significant considering the fact that many systems exhibiting
novel electronic properties can be studied efficiently using
samples in polycrystalline forms, particulary where it is
difficult to prepare single crystals.

The O 1$s$ spectra in Ca substituted compositions exhibit two peak
structure, which can be attributed to the difference in Madelung
potential at different oxygen sites in the structure. Decrease in
temperature leads to a significant modification in the local
structure in the Ca dominated compositions. While this provides
one possible explanation of the proximity of these compounds to
the quantum criticality, it is highly desirable to investigate the
issue further. We hope that these results would help to initiate
further studies ($x$-ray diffraction, extended $x$-ray fine
structure {\it etc.}) in this direction.


\begin{thebibliography}{99}
%
\bibitem{millis} A.J. Millis, Nature 392 (1998) 147.
%
\bibitem{mannella} N. Mannella {\em et al.}, Nature 438 (2005) 474.
%
\bibitem{ruth-nfl} K. Maiti, R.S. Singh, V.R.R. Medicherla,
(unpublished); Cond-mat/0604648.
%
%
\bibitem{rss} R.S. Singh, P.L. Paulose, K. Maiti, Solid State Physics (India), 49 (2004)
876.
%
\bibitem{cao} G. Cao, S. McCall, M. Shepard, J.E. Crow, R.P.
Guertin, Phys. Rev. B 56 (1997) 321.
%
\bibitem{nfl} P. Khalifah, I. Ohkubo, H. Christen, D. Mandrus, Phys.
Rev. B 70 (2004) 134426; Y. S. Lee, Jaejun Yu, J. S. Lee, T. W.
Noh, T.-H. Gimm, Han-Yong Choi, C. B. Eom, Phys. Rev. B 66 (2002)
041104(R).
%
\bibitem{klein} L. Klein, L. Antognazza, T.H. Geballe, M.R. Beasley,
A. Kapitulnik, Phys. Rev. B 60 (1999) 1448.
%
\bibitem{ruth-corr} K. Maiti, R.S. Singh, Phys. Rev. B 71 (2005) 161102(R).
%
\bibitem{ruth-band} K. Maiti, Phys. Rev. B 73 (2006) 235110;
Cond-mat/0605553.
%
\bibitem{don-sun} S. Doniach, M. \v{S}unji\'{c}, J. Phys. C: Solid
State Phys. 3 (1970) 285.
%
%
\bibitem{wien} P. Blaha, K. Schwarz, G.K.H. Madsen, D. Kvasnicka, J. Luitz, {\bf
WIEN2k}, An Augmented Plane Wave + Local Orbitals Program for
Calculating Crystal Properties, Karlheinz Schwarz, Techn.
Universit\"{a}t Wien, Austria, 2001, (ISBN 3-9501031-1-2).
%
\bibitem{ruth-epl} R.S. Singh, K. Maiti, (unpublished) (2006); Cond-mat/0605552.
%
\end{thebibliography}
\end{document}